# Kinetic Energy Matrix Elements for a two-electron Atom with Extended Hylleraas-CI Wave Function

B PADHY

Former Faculty Member,
Department of Physics, Khallikote (Autonomous) College,
Brahmapur-760001,Odisha.
Email: bholanath.padhy@gmail.com

**Abstract :** Some typical kinetic energy integrals which arise in the application of extended Hylleraas-configuration interaction (E-Hy-CI) function in the framework of Rayleigh-Ritz method of variation, have been evaluated analytically for two-electron atomic systems. Closed-form expressions for the corresponding integrals which occur in the application of Hylleraas-CI functions, have also been derived as special cases.

## 1. Introduction

For highly precise computation of electronic structure in multielectron atoms employing nonrelativistic quantum mechanics in the framework of Rayleigh-Ritz variational procedure, one needs to take into account the correlation between various electrons in the atom. There are basically two methods of calculation which include these electron-electron correlations, namely, (i) Hylleraas (Hy) method [1], and (ii) configuration-interaction (CI) method. Various Hy and CI studies have been reported in a review article by Silverman and Brigman [2]. It is known that CI method is plagued with the weakness of extremely slow convergence. Though Hy method is regarded as the most powerful method among the existing theoretical approaches to produce results of high accuracy [3], one faces great computational difficulty when it is applied for calculations for atoms with a large number of electrons (N>4).

Subsequently, an alternative method, known as Hylleraas-CI (Hy-CI) method, was systematically developed by Sims and Hagstrom [4] in early seventies by hybridization of Hy and CI methods. This method has been widely applied successfully over the last few decades for calculations even for atoms with a large number of electrons. For knowledge about the progressive development of Hy-CI method and its applications, the reader is advised to go through the representative papers for potential energy calculation [5] and for kinetic energy calculation [6], and references therein.

It is expected that still quicker convergence will be achieved with smaller expansion lengths in the wave function expansion with the involvement of exponential correlation in Hy and Hy-CI methods. Accordingly, Hy-CI method has been extended further by including exponential correlation, and identified as Extended Hy-CI (E-Hy-CI) method [7]. There are few other recent reports in the literature wherein overlap and potential energy integrals involving exponential correlation with 'unlinked' $r_{ij}$'s, and spherically symmetric atomic $s$ Slater-type orbitals (STO's) have been evaluated in closed-form [8-11]. Also such integrals involving exponential correlation and nonspherically symmetric atomic STO's have already been evaluated analytically [12].

Very recently, Harris, in a series of papers [13-15], has outlined a method for evaluating kinetic energy matrix elements for Hy-CI and E-Hy-CI functions analytically making use of properties of vector spherical harmonics. The method of evaluation is rather simpler compared to that used in [6,16,17].

In this paper a basic kinetic energy integral, which arises in the application of E-Hy-CI method, and denoted as $K^{EHCI}$, is analytically evaluated for a two-electron atom. Closed-form expressions for several other kinetic energy integrals are obtained from the expression for $K^{EHCI}$ by employing the method of parametric differentiation and taking certain limits as per requirement.

The plan of this paper is as follows. In section 2, all the integrals to be evaluated are clearly defined. In section 3, first the basic kinetic energy integral $K^{EHCI}$ is reduced to a simplified form to express it easily in terms of overlap/potential energy integrals. Then all the other kinetic energy integrals corresponding to E-Hy-CI method are reduced to simplified forms by parametric differentiations of $K^{EHCI}$. Closed-form expression for the basic integral $K^{EHCI}$ is obtained in section 4, and then other kinetic energy integrals which arise in E-Hy-CI method are analytically evaluated. In section 5, simplified integrable expressions for the corresponding kinetic energy integrals which arise in Hy-CI method are obtained as special cases from those integrals of E-Hy-CI method by taking certain limits as required, and then these integrals are also easily evaluated analytically. Further, one kinetic energy integral arising in CI method is also analytically evaluated. In the last section, some concluding remarks are given.

## 2. Definition of various two-electron atomic integrals

Let $\vec{r}_1$ and $\vec{r}_2$ be the position vectors of the two electrons with respect to the nucleus in the infinite nuclear mass approximation. Obviously, $r_{12} = |\vec{r}_1 - \vec{r}_2|$ is the distance between the two electrons. With spherical polar coordinates, one can write $\vec{r}$ as $\vec{r} \equiv (r, \theta, \varphi)$. Thus the following set of unnormalised atomic STO's is taken as the orbital basis:

$$\varphi_a(j) = r_j^{n_a-1} e^{-\alpha_a r_j} Y_{l_a}^{m_a}(j), \ \alpha_a > 0 , \tag{1}$$

where $Y_l^m$ is an orthonormal spherical harmonic with its arguments, here denoted as $j$, being the angular coordinates of $\vec{r}_j$, and the subscript '$a$' signifies a particular STO. In the Condon and Shortley phase convention, $Y_l^m$ is defined in [4,13], where $l$ and $m$ quantum numbers give the order and degree of the spherical harmonic, with $n-1 \geq l \geq 0$, $n$ being the radial quantum number.

### 2.1 Overlap / potential energy integrals

The most general type of overlap/potential energy integrals which arise in calculations with E-Hy-CI functions for a two-electron atomic system are of the form

$$I^{EHCI} = \langle \varphi_a(1) \varphi_b(2) | R | \varphi_d(1) \varphi_e(2) \rangle, \tag{2}$$

where $R$ is an operator given by

$$R = r_{12}^\nu e^{-\beta r_{12}}, \ \nu \geq -1, \ \beta \geq 0. \tag{3}$$

It is to be noted that if $\beta = 0$ and $\nu \geq 1$ simultaneously, one gets integrals corresponding to Hy-CI functions without involving exponential correlation. If simultaneously $\beta = 0$ and $\nu = 0$, then the integrals corresponding to CI function are obtained as special cases.

### 2.2 Kinetic energy with CI functions

If only CI method is considered, kinetic energy integrals are defined by

$$K^{CI} = \left\langle \varphi_a(1)\,\varphi_b(2) \left| -\tfrac{1}{2}\vec{\nabla}_1^2 \right| \varphi_d(1)\,\varphi_e(2) \right\rangle. \tag{4}$$

### 2.3 Kinetic energy integrals with Hy-CI functions

The following types of kinetic energy integrals arise while employing the Hy-CI method:

$$K_1^{HCI} = \left\langle r_{12}\,\varphi_a(1)\varphi_b(2) \left| -\tfrac{1}{2}\vec{\nabla}_1^2 \right| \varphi_d(1)\varphi_e(2) \right\rangle, \tag{5}$$

$$K_2^{HCI} = \left\langle \varphi_a(1)\varphi_b(2) \left| -\tfrac{1}{2}\vec{\nabla}_1^2 \right| \varphi_d(1)\varphi_e(2)\,r_{12} \right\rangle, \tag{6}$$

$$K_3^{HCI} = \left\langle r_{12}\,\varphi_a(1)\varphi_b(2) \left| -\tfrac{1}{2}\vec{\nabla}_1^2 \right| \varphi_d(1)\varphi_e(2)\,r_{12} \right\rangle. \tag{7}$$

It is to be mentioned here that, since the kinetic energy operator is Hermitian, $K_2^{HCI}$ can also be written as

$$K_2^{HCI} = \left\langle \varphi_d(1)\varphi_e(2)\,r_{12} \left| -\tfrac{1}{2}\vec{\nabla}_1^2 \right| \varphi_a(1)\varphi_b(2) \right\rangle^*, \tag{8}$$

where the superscript symbol '*' stands for the complex conjugate of the integral.

### 2.4 Kinetic energy integrals in E-Hy-CI method

For doing calculations employing E-Hy-CI method, the following different types of kinetic energy integrals are to be evaluated.

$$K^{EHCI}(w,w') = \left\langle \varphi_a(1)\varphi_b(2)e^{-wr_{12}} \left| -\tfrac{1}{2}\vec{\nabla}_1^2 \right| \varphi_d(1)\varphi_e(2)e^{-w'r_{12}} \right\rangle, \tag{9}$$

$$K_1^{EHCI}(w,w') = \left\langle r_{12}\,\varphi_a(1)\varphi_b(2)e^{-wr_{12}} \left| -\tfrac{1}{2}\vec{\nabla}_1^2 \right| \varphi_d(1)\varphi_e(2)e^{-w'r_{12}} \right\rangle, \tag{10}$$

$$K_2^{EHCI}(w,w') = \left\langle \varphi_a(1)\varphi_b(2)e^{-wr_{12}} \left| -\tfrac{1}{2}\vec{\nabla}_1^2 \right| \varphi_d(1)\varphi_e(2)r_{12}e^{-w'r_{12}} \right\rangle, \tag{11}$$

$$K_3^{EHCI}(w,w') = \left\langle \varphi_a(1)\varphi_b(2)r_{12}e^{-wr_{12}} \left| -\tfrac{1}{2}\vec{\nabla}_1^2 \right| \varphi_d(1)\varphi_e(2)r_{12}e^{-w'r_{12}} \right\rangle. \tag{12}$$

Here $w$ and $w'$ are exponential parameters, which may or may not be equal depending on the two-electron states. Also $w, w' \geq 0$.

## 3. Simplification of the integrals

By employing Kolos-Roothan transformation [18], the right hand side expression in Eq. (9) is simplified to obtain the following easily integrable expression for the integral $K^{EHCI}(w, w')$:

$$K^{EHCI}(w,w') = \frac{1}{2} \int d\vec{r}_2 \, \varphi_b^*(2) \, \varphi_e(2) \int d\vec{r}_1 \, e^{-(w+w')r_{12}}$$

$$\times \left[ \varphi_a^*(1) \varphi_d(1) ww' - \frac{w'}{w+w'} \varphi_d(1) \left( \vec{\nabla}_1^2 \varphi_a^*(1) \right) \right.$$

$$\left. - \frac{w}{w+w'} \varphi_a^*(1) \left( \vec{\nabla}_1^2 \varphi_d(1) \right) \right]. \tag{13}$$

Looking at the right hand side expressions in Eqs. (9-12), it is clearly observed that Eqs. (10-12) can be generated from Eq.(9) by parametric differentiation with respect to $w$ or $w'$ or both as per requirement. Accordingly, easily integrable simplified expressions for integrals in Eqs. (10-12) can be obtained from Eq. (13) by parametric differentiation. Thus

$$K_1^{EHCI}(w,w') = \left( -\frac{\partial}{\partial w} \right) K^{EHCI}(w,w')$$

$$= \frac{1}{2} \int d\vec{r}_2 \, \varphi_b^*(2) \, \varphi_e(2) \int d\vec{r}_1 \, e^{-(w+w')r_{12}}$$

$$\times \left[ w'(wr_{12} - 1)\varphi_a^*(1)\varphi_d(1) - \left\{ \frac{w'r_{12}}{w+w'} + \frac{w'}{(w+w')^2} \right\} \varphi_d(1) \vec{\nabla}_1^2 \varphi_a^*(1) \right.$$

$$\left. + \left\{ \frac{w'}{(w+w')^2} - \frac{wr_{12}}{w+w'} \right\} \varphi_a^*(1) \vec{\nabla}_1^2 \varphi_d(1) \right], \tag{14}$$

$$K_2^{EHCI}(w,w') = \left(-\frac{\partial}{\partial w'}\right) K^{EHCI}(w,w')$$

$$= \frac{1}{2}\int d\vec{r}_2\, \varphi_b^*(2)\, \varphi_e(2) \int d\vec{r}_1\, e^{-(w+w')r_{12}}$$

$$\times \left[ w(w'r_{12}-1)\varphi_a^*(1)\varphi_d(1) + \left\{\frac{w}{(w+w')^2} - \frac{w'r_{12}}{w+w'}\right\}\varphi_d(1)\vec{\nabla}_1^2\varphi_a^*(1) \right.$$

$$\left. -\left\{\frac{w}{(w+w')^2} + \frac{wr_{12}}{w+w'}\right\}\varphi_a^*(1)\vec{\nabla}_1^2\varphi_d(1) \right]. \qquad (15)$$

It is easy to observe that the integrable expression for $K_3^{HCI}$ can be obtained in either of the following two ways:

$$K_3^{EHCI}(w,w') = \left(-\frac{\partial}{\partial w}\right)\left(-\frac{\partial}{\partial w'}\right) K^{EHCI}(w,w') = \left(-\frac{\partial}{\partial w}\right) K_2^{EHCI}(w,w'), \qquad (16)$$

$$K_3^{EHCI}(w,w') = \left(-\frac{\partial}{\partial w'}\right)\left(-\frac{\partial}{\partial w}\right) K^{EHCI}(w,w') = \left(-\frac{\partial}{\partial w'}\right) K_1^{EHCI}(w,w'). \qquad (17)$$

Performing the differentiations as per Eqs. (16) and (17), the same expression is obtained for $K_3^{EHCI}$, as expected. Thus

$$K_3^{EHCI}(w,w') = \frac{1}{2}\int d\vec{r}_2\, \varphi_b^*(2)\varphi_e(2) \int d\vec{r}_1\, e^{-(w+w')r_{12}}$$

$$\times \left[ \left\{1-(w'+w)r_{12} + ww'r_{12}^2\right\}\varphi_a^*(1)\,\varphi_d(1) \right.$$

$$\left. + \left\{\frac{w'-w}{(w'+w)^3} + \frac{w'-w}{(w'+w)^2}r_{12} - \frac{w}{w+w'}r_{12}^2\right\}\varphi_a^*(1)\vec{\nabla}_1^2\varphi_d(1) \right.$$

$$-\left\{\frac{w'-w}{(w'+w)^3}+\frac{w'-w}{(w'+w)^2}r_{12}+\frac{w'}{w+w'}r_{12}^2\right\}\varphi_d(1)\vec{\nabla}_1^2\varphi_a^*(1)\Bigg]. \quad (18)$$

## 4. Analytic evaluation of integrals in E-Hy-CI method

Expressing $\vec{\nabla}^2$ in spherical polar coordinates, it is straightforward to derive the following equations [13]:

$$-\frac{1}{2}\vec{\nabla}_1^2\varphi_a^*(1)=\left[\frac{(l_a+n_a)(l_a-n_a+1)}{2r_1^2}+\frac{n_a\alpha_a}{r_1}-\frac{\alpha_a^2}{2}\right]\varphi_a^*(1), \quad (19)$$

$$-\frac{1}{2}\vec{\nabla}_1^2\varphi_d(1)=\left[\frac{(l_d+n_d)(l_d-n_d+1)}{2r_1^2}+\frac{n_d\alpha_d}{r_1}-\frac{\alpha_d^2}{2}\right]\varphi_d(1). \quad (20)$$

Inserting Eqs. (19) and (20) in Eqs. (13-15) and (18), each of the integrals $K^{EHCI}, K_1^{EHCI}, K_2^{EHCI}$ and $K_3^{EHCI}$ can be expressed as a combination of several overlap/potential energy integrals involving exponential correlation as per Eq. (2) and evaluated analytically. Thus from Eq. (13)

$$K^{EHCI}(w,w')=\frac{1}{2}\left[ww'-\frac{w'}{w+w'}\alpha_a^2-\frac{w}{w+w'}\alpha_d^2\right]\langle S\rangle$$

$$+\frac{1}{2}\left[\frac{w'}{w+w'}(l_a+n_a)(l_a-n_a+1)+\frac{w}{w+w'}(l_d+n_d)(l_d-n_d+1)\right]\langle S/r_1^2\rangle$$

$$+\left[\frac{w'}{w+w'}n_a\alpha_a+\frac{w}{w+w'}n_d\alpha_d\right]\langle S/r_1\rangle, \quad (21)$$

where

$$S=e^{-(w+w')r_{12}}, \quad (22)$$

and the symbol $\langle T\rangle$, in general, stands for the expression given by

$$\langle T \rangle = \langle \varphi_a(1)\, \varphi_b(2) | T | \varphi_d(1)\, \varphi_e(2) \rangle. \tag{23}$$

Similarly, from Eq. (14), the following expression for $K_1^{EHCI}(w, w')$ is obtained:

$$K_1^{EHCI}(w, w') = \frac{1}{2}\left[ ww' - \frac{w'}{w+w'} \alpha_a^2 - \frac{w}{w+w'} \alpha_d^2 \right] \langle S\, r_{12} \rangle$$

$$+ \frac{1}{2}\left[ -w' + \frac{w'}{(w+w')^2}(\alpha_d^2 - \alpha_a^2) \right] \langle S \rangle$$

$$+ \frac{1}{2}\frac{w'}{(w+w')^2}\left[ (l_a + n_a)(l_a - n_a + 1) - (l_d + n_d)(l_d - n_d + 1) \right] \langle S/r_1^2 \rangle$$

$$+ \frac{w'}{(w+w')^2}(n_a \alpha_a - n_d \alpha_d) \langle S/r_1 \rangle$$

$$+ \frac{1}{2}\left[ \frac{w'}{w+w'}(l_a + n_a)(l_a - n_a + 1) + \frac{w}{w+w'}(l_d + n_d)(l_d - n_d + 1) \right] \langle S\, r_{12}/r_1^2 \rangle$$

$$+ \left[ \frac{w'}{w+w'} n_a \alpha_a + \frac{w}{w+w'} n_d \alpha_d \right] \langle S\, r_{12}/r_1 \rangle, \tag{24}$$

where $S$ is given by Eq. (22).

Starting with Eq.(15) and proceeding in the same manner, one can obtain

$$K_2^{EHCI}(w, w') = \frac{1}{2}\left[ ww' - \frac{w'}{w+w'} \alpha_a^2 - \frac{w}{w+w'} \alpha_d^2 \right] \langle S\, r_{12} \rangle$$

$$+ \frac{1}{2}\left[ \frac{w}{(w+w')^2}(\alpha_a^2 - \alpha_d^2) - w \right] \langle S \rangle$$

$$+ \frac{1}{2}\frac{w'}{(w+w')^2}\left[ (l_d + n_d)(l_d - n_d + 1) - (l_a + n_a)(l_a - n_a + 1) \right] \langle S/r_1^2 \rangle$$

$$+ \frac{w}{(w+w')^2}(n_d\alpha_d - n_a\alpha_a)\langle S/r_1 \rangle$$

$$+ \frac{1}{2}\left[\frac{w'}{w+w'}(l_a+n_a)(l_a-n_a+1) + \frac{w}{w+w'}(l_d+n_d)(l_d-n_d+1)\right]\langle S r_{12}/r_1^2 \rangle$$

$$+ \left[\frac{w'}{w+w'}n_a\alpha_a + \frac{w}{w+w'}n_d\alpha_d\right]\langle S r_{12}/r_1 \rangle. \tag{25}$$

Expression in Eq. (18) is further simplified similarly to get

$$K_3^{EHCI}(w,w') = \frac{1}{2}\left[1 + \frac{w'-w}{(w+w')^3}(\alpha_d^2 - \alpha_a^2)\right]\langle S \rangle$$

$$+ \frac{1}{2}\left[-(w+w') + \frac{w'-w}{(w+w')^2}(\alpha_d^2 - \alpha_a^2)\right]\langle S r_{12} \rangle$$

$$+ \frac{1}{2}\left[ww' - \frac{w'}{w+w'}\alpha_a^2 - \frac{w}{w+w'}\alpha_d^2\right]\langle S r_{12}^2 \rangle$$

$$+ \frac{1}{2}\frac{w'-w}{(w'+w)^3}\left[(l_a+n_a)(l_a-n_a+1) - (l_d+n_d)(l_d-n_d+1)\right]\langle S/r_1^2 \rangle$$

$$+ \frac{w'-w}{(w+w')^3}(n_a\alpha_a - n_d\alpha_d)\langle S/r_1 \rangle$$

$$+ \frac{1}{2}\frac{w'-w}{(w'+w)^2}\left[(l_a+n_a)(l_a-n_a+1) - (l_d+n_d)(l_d-n_d+1)\right]\langle S r_{12}/r_1^2 \rangle$$

$$+ \frac{w'-w}{(w+w')^2}(n_a\alpha_a - n_d\alpha_d)\langle S r_{12}/r_1 \rangle$$

$$+ \frac{1}{2}\left[\frac{w'}{w+w'}(l_a+n_a)(l_a-n_a+1) + \frac{w}{w+w'}(l_d+n_d)(l_d-n_d+1)\right]\langle S r_{12}^2/r_1^2 \rangle$$

$$+\left[\frac{w'}{w'+w}n_a\alpha_a + \frac{w}{w+w'}n_d\alpha_d\right]\langle S\, r_{12}^2/r_1\rangle. \tag{26}$$

## 5. Simplification and analytic evaluation of integrals in CI and Hy-CI methods

(i) The integral $K^{CI}$ as defined in Eq. (4) can be analytically evaluated employing Eq. (20). Thus

$$K^{CI} = \frac{1}{2}(l_d + n_d)(l_d - n_d + 1)\langle\varphi_a(1)\varphi_b(2)|1/r_1^2|\varphi_d(1)\varphi_e(2)\rangle$$

$$+ n_d\alpha_d \langle\varphi_a(1)\varphi_b(2)|1/r_1|\varphi_d(1)\varphi_e(2)\rangle$$

$$- \frac{1}{2}\alpha_d^2 \langle\varphi_a(1)\varphi_b(2)|\varphi_d(1)\varphi_e(2)\rangle. \tag{27}$$

(ii) Inserting Eq. (20) in Eq. (5), the following closed-form expression for the integral $K_1^{HCI}$ is obtained:

$$K_1^{HCI} = \frac{1}{2}(l_d + n_d)(l_d - n_d + 1)\langle\varphi_a(1)\varphi_b(2)|r_{12}/r_1^2|\varphi_d(1)\varphi_e(2)\rangle$$

$$+ n_d\alpha_d \langle\varphi_a(1)\varphi_b(2)|r_{12}/r_1|\varphi_d(1)\varphi_e(2)\rangle$$

$$- \frac{1}{2}\alpha_d^2 \langle\varphi_a(1)\varphi_b(2)|r_{12}|\varphi_d(1)\varphi_e(2)\rangle, \tag{28}$$

which is exactly identical with Eq.(12) in [13].

(iii) The integral $K_2^{HCI}$ as defined in Eq. (6) can be analytically evaluated via Eq. (8), making use of Eq. (19). Thus

$$K_2^{HCI} = \frac{1}{2}(l_a + n_a)(l_a - n_a + 1)\langle\varphi_a(1)\varphi_b(2)|r_{12}/r_1^2|\varphi_d(1)\varphi_e(2)\rangle$$

$$+ n_a\alpha_a \langle\varphi_a(1)\varphi_b(2)|r_{12}/r_1|\varphi_d(1)\varphi_e(2)\rangle$$

$$- \frac{1}{2}\alpha_a^2 \langle\varphi_a(1)\varphi_b(2)|r_{12}|\varphi_d(1)\varphi_e(2)\rangle. \tag{29}$$

It is observed that the right hand side expressions in Eqs. (28) and (29) are not identical, in general. Thus the kinetic energy integrals defined in Eqs. (5) and (6) are not identical, in general.

(iv) To evaluate the integral $K_3^{HCI}$ as defined in Eq. (7), one looks at the Eq. (12) which defines the integral $K_3^{EHCI}(w, w')$. It is observed that with $w = w' = 0$, Eq. (12) reduces to Eq. (7). Hence a simplified expression for $K_3^{HCI}$ can be obtained from that of $K_3^{EHCI}$ by setting first $w' = w$ and then letting $w \to 0$ in Eq. (18). Thus

$$K_3^{HCI} = \frac{1}{2} \int d\vec{r}_2\, \varphi_b^*(2)\, \varphi_e(2) \int d\vec{r}_1$$

$$\times \left[ \varphi_a^*(1)\varphi_d(1) - \frac{1}{2} r_{12}^2 \varphi_a^*(1)\, \vec{\nabla}_1^2 \varphi_d(1) - \frac{1}{2} r_{12}^2 \varphi_d(1)\, \vec{\nabla}_1^2 \varphi_a^*(1) \right]. \quad (30)$$

The case $w' = w = 0$ has been discussed earlier by Kolos and Roothan [18]. Next inserting Eqs. (19) and (20) in Eq. (30), the following closed-form expression for $K_3^{HCI}$ is obtained:

$$K_3^{HCI} = \frac{1}{2} \langle \varphi_a(1)\varphi_b(2) \mid \varphi_d(1)\varphi_e(2) \rangle$$

$$+ \frac{1}{4}[(l_a + n_a)(l_a - n_a + 1) + (l_d + n_d)(l_d - n_d + 1)]$$

$$\times \langle \varphi_a(1)\, \varphi_b(2) \mid r_{12}^2 / r_1^2 \mid \varphi_d(1)\varphi_e(2) \rangle$$

$$+ \frac{1}{2}(n_a \alpha_a + n_d d_d)\, \langle \varphi_a(1)\varphi_b(2) \mid r_{12}^2 / r_1 \mid \varphi_d(1)\varphi_e(2) \rangle$$

$$- \frac{1}{4}\left( \alpha_a^2 + \alpha_d^2 \right) \langle \varphi_a(1)\varphi_b(2) \mid r_{12}^2 \mid \varphi_d(1)\varphi_e(2) \rangle. \quad (31)$$

This expression in Eq. (31) can be compared with the right hand side expression in Eq. (27) in [13], which has been derived for the same integral by

Harris in a different approach by making use of vector spherical harmonics, and later pointing out a minor misprint.

The validity of the derivation leading to Eq. (31) will be established by calculating the values of the expression given here and that reported by Harris [13].

## 6. Conclusion

All the kinetic energy integrals defined here for a two-electron atomic system have been expressed in terms of the overlap/potential energy integrals which can be easily evaluated analytically. Thus, closed-form expression is obtained for the total energy of the system. The energy of various states of such a system can be estimated by applying the method of variation, and choosing the expansion length of the wave function proposed. The numerical results can be compared with energy values of few states of a helium atom reported recently [19].

## Acknowledgements

The author is extremely grateful to Dr. J. S. Sims and Dr. M. B. Ruiz for providing periodic information as and when required throughout this work. Also, I am highly indebted to Prof. N. Barik for several useful discussions relating to this work, and to Prof. S. K. Patra for critically going through the manuscript.